\documentclass[journal,10pt,doublecolumn]{IEEEtran}

\usepackage{setspace}
\usepackage{amsmath}
\usepackage{graphicx}
\usepackage{cite}
\usepackage{amsfonts}
\usepackage{amsthm}
\usepackage{algorithm}
\usepackage{algorithmic}
\usepackage{amssymb}
\ifCLASSOPTIONcompsoc
\usepackage[caption=false,font=normalsize,labelfont=sf,textfont=sf]{subfig}
\else
\usepackage[caption=false,font=footnotesize]{subfig}
\fi

\theoremstyle{definition} 

\begin{document}

\title{Secure Wireless Communication via Intelligent Reflecting Surface}

\author{Miao~Cui, Guangchi~Zhang,~\IEEEmembership{Member,~IEEE,} and Rui~Zhang,~\IEEEmembership{Fellow,~IEEE}
\thanks{This work was supported in part by the National Natural Science Foundation of China under Grant 61571138, in part by the Science and Technology Plan Project of Guangdong Province under Grants 2017B090909006, 2018A050506015, and 2019B010119001, and in part by the Science and Technology Plan Project of Guangzhou City under Grants 201803030028 and 201904010371.
	
M. Cui and G. Zhang are with the School of Information Engineering, Guangdong University of Technology, Guangzhou, China (email: \{cuimiao, gczhang\}@gdut.edu.cn). R. Zhang is with the Department of Electrical and Computer Engineering, National University of Singapore, Singapore 117576 (email: elezhang@nus.edu.sg). G. Zhang is the corresponding author.}   }

\maketitle

\vspace{-0.5cm}

\begin{abstract}
An intelligent reflecting surface (IRS) can adaptively adjust the phase shifts of its reflecting units to strengthen the desired signal and/or suppress the undesired signal. In this letter, we investigate an IRS-aided secure wireless communication system where a multi-antenna access point (AP) sends confidential messages to a single-antenna user in the presence of a single-antenna eavesdropper. In particular, we consider the challenging scenario where the eavesdropping channel is stronger than the legitimate communication channel and they are also highly correlated in space. We maximize the secrecy rate of the legitimate communication link by jointly designing the AP's transmit beamforming and the IRS's reflect beamforming. While the resultant optimization problem is difficult to solve, we propose an efficient algorithm to obtain high-quality suboptimal solution for it by applying the alternating optimization and semidefinite relaxation methods. Simulation results show that the proposed design significantly improves the secrecy communication rate for the considered setup over the case without using the IRS, and outperforms a heuristic scheme.
\end{abstract}


\IEEEpeerreviewmaketitle

\vspace{-0.4cm}
\section{Introduction}
With the advancement in micro electromechanical systems (MEMS) and metamaterial techniques, it becomes feasible to control the phase shift of reflected signal in real time via a programmable surface. This has enabled a new wireless device, named intelligent reflecting surface (IRS), which can be flexibly deployed in wireless networks to improve their performance in various ways \cite{Wu2019}. An IRS is usually composed of a large number of low-cost, passive, reflecting units, each being able to reflect the incident wireless signal with an adjustable phase shift \cite{Cui2014}. By adaptively tuning the phase shifts of the reflecting units, the signal reflected by IRS can add constructively or destructively with the non-IRS-reflected signal at the receiver to enhance the desired signal or suppress the undesired signal such as interference \cite{Tan2016}. In an IRS-aided communication system, active transmit beamforming at the transmitter and passive reflect beamforming at the IRS can be jointly designed to improve the performance, e.g., transmit power minimization \cite{Wu2018, Wu2018a, Wu2018b} and energy efficiency maximization \cite{Huang2018}. It is worth noting that as compared to IRS, there is another technology called large intelligent surface (LIS) \cite{Hu2018}, which can operate in either active or passive mode, while it resembles IRS in the passive mode.

\begin{figure}[!t]
	\centering
	\includegraphics[width=0.4\textwidth]{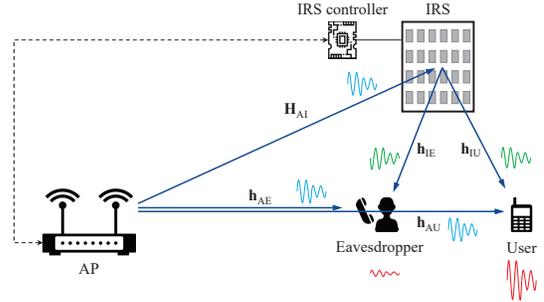}  \vspace{-0.2cm}
	\caption{IRS-aided secure communication from an AP to a user in the presence of an eavesdropper.}
	\label{FigSystem} \vspace{-0.6cm}
\end{figure}

On the other hand, physical-layer security has been thoroughly investigated in wireless communications. To maximize secrecy communication rate, various techniques such as jamming with artificial noise (AN) and multi-antenna beamforming have been proposed \cite{Khisti2010, Mukherjee2014, Liu2014}. However, in scenarios where the channel of the legitimate communication link and that of the eavesdropping link are spatially highly correlated and the average power of the former is weaker than that of the latter as shown in Fig. \ref{FigSystem}, the achievable secrecy rate is very limited, even with the aforementioned techniques applied.

In this letter, we apply IRS to tackle the above challenge. As shown in Fig. \ref{FigSystem}, we consider the secure communication from a multi-antenna access point (AP) to a single-antenna user in the presence of a single-antenna eavesdropper, where an IRS is deployed in the vicinity of the user and the eavesdropper. First, by adjusting the phase shifts of the IRS's reflecting units, the reflected signal by the IRS is added constructively with the non-reflected signal at the user to boost its received signal power, while being destructively added with that at the eavesdropper to cancel its received signal, both leading to improved secrecy rate for the user. In addition, the AP's transmit beamforming can be designed to strike a balance between the signal power beamed towards the IRS and that to the user/eavesdropper for signal enhancement/cancellation, respectively. Thus, by jointly optimizing the active transmit beamforming at the AP and the passive reflect beamforming at the IRS, the secrecy rate for the user can be maximized. However, this optimization problem is difficult to solve due to its non-convexity and coupled variables. To tackle this problem, we propose an efficient algorithm to solve it approximately, based on the alternating optimization, semidefinite relaxation (SDR) and Gaussian randomization methods. Simulation results show the significant gains in secrecy rate by the proposed design for the considered setup, as compared to the case without using the IRS as well as a heuristic beamforming design.

\vspace{-0.3cm}
\section{System Model and Problem Formulation}
As shown in Fig. \ref{FigSystem}, we consider that an AP with $M$ antennas communicates with a single-antenna user in the presence of a single-antenna eavesdropper. An IRS with $N$ reflecting units is deployed to assist in the secure communication from the AP to the user. The IRS is equipped with a controller, which coordinates the AP and IRS for both channel acquisition and data transmission \cite{Wu2018}. All channels in the system are assumed to experience quasi-static flat-fading. The channel coefficients from the AP to the IRS, from the AP to the user, from the AP to the eavesdropper, from the IRS to the user, and from the IRS to the eavesdropper are denoted by $\mathbf{H}_{\text{AI}} \in \mathbb{C}^{ N \times M}$, $\mathbf{h}_{\text{AU}}  \in \mathbb{C}^{1 \times M}$, $\mathbf{h}_{\text{AE}} \in \mathbb{C}^{1 \times M} $, $\mathbf{h}_{\text{IU}} \in \mathbb{C}^{1 \times N} $, and $\mathbf{h}_{\text{IE}} \in \mathbb{C}^{1 \times N} $, respectively. Here, $\mathbb{C}^{m \times n}$ denotes the set of $m \times n$ complex-valued matrices. To characterize the performance limit of the considered IRS-aided secrecy communication system, we assume that the global channel state information (CSI) on all the above involved channels is perfectly known at the AP/IRS for their joint design of transmit/reflect beamforming. Note that it is reasonable to assume that $\mathbf{h}_{\text{AE}}$ and $\mathbf{h}_{\text{IE}}$ are known in the scenario when the eavesdropper is an active user in the system but untrusted by the legitimate receiver \cite{Mukherjee2014}.

The AP transmits confidential message $s$ with zero mean and unit variance to the user via beamforming. The beamforming vector is denoted by $\mathbf{w} \in \mathbb{C}^{M \times 1}$, which satisfies the following constraint
\begin{equation}
\| \mathbf{w} \|^2  \leq P_{\text{AP}},
\end{equation}
where $\| \cdot \|$ denotes the Euclidean norm, and $P_{\text{AP}}$ is the maximum transmit power of the AP. Each unit of the IRS reflects the received signal from the AP with an adjustable phase shift. We model the reflection by the units of the IRS using $\mathbf{q} \triangleq [ q_1, \ldots, q_N ]^T$, where $q_n = \beta_n e^{j \theta_n}$, and $\theta_n \in [0, 2\pi)$ and $\beta_n \in [0,1]$, $n=1,\ldots,N$, denote the phase shift and amplitude reflection coefficient of the $n$th unit, respectively. Here, the superscript $T$ denotes the transpose operation. For simplicity, we set $\beta_n = 1$, $\forall n$, to achieve the maximum reflecting power gain, thus $\mathbf{q}$ should satisfy
\begin{equation}
| q_n |=  1,  \; \forall n.  
\end{equation}

\vspace{-0.3cm}
Neglecting the signals reflected by the IRS two or more times due to severe path loss, the received signals at the user and the eavesdropper can be respectively expressed as
\begin{equation}
    y_{\text{U}} =  ( \mathbf{h}_{\text{IU}} \mathbf{Q} \mathbf{H}_{\text{AI}} + \mathbf{h}_{\text{AU}} ) \mathbf{w} s + n_{\text{U}},
\end{equation}
\begin{equation}
y_{\text{E}} = ( \mathbf{h}_{\text{IE}} \mathbf{Q} \mathbf{H}_{\text{AI}} + \mathbf{h}_{\text{AE}} ) \mathbf{w} s   + n_{\text{E}},
\end{equation}
where $\mathbf{Q} \triangleq \text{diag}(\mathbf{q})$ denotes a diagonal matrix whose diagonal elements are the corresponding elements of the vector $\mathbf{q}$, and $n_{\text{U}}$ and $n_{\text{E}}$ denote the Gaussian noises at the user and the eavesdropper with mean zero and variances $\sigma_{\text{U}}^2$ and $\sigma_{\text{E}}^2$, respectively. Thus, the secrecy rate from the AP to the user in bits/second/Hertz (bps/Hz) can be expressed as \cite{Khisti2010}
\begin{equation}   \label{EquRsec}
R_{\text{sec}} = [ R_{\text{U}} - R_{\text{E}} ]^+,
\end{equation}
where $[z]^+ \triangleq \max(z,0)$, and
\begin{equation} \label{EquRU}
R_{\text{U}} = \log_2 \left( 1 + \frac{ | ( \mathbf{h}_{\text{IU}} \mathbf{Q} \mathbf{H}_{\text{AI}} + \mathbf{h}_{\text{AU}} ) \mathbf{w} |^2 }{ \sigma_{\text{U}}^2 }  \right),
\end{equation}
\begin{equation}  \label{EquRE}
R_{\text{E}} = \log_2 \left( 1 + \frac{ | ( \mathbf{h}_{\text{IE}} \mathbf{Q} \mathbf{H}_{\text{AI}} + \mathbf{h}_{\text{AE}} ) \mathbf{w} |^2 }{ \sigma_{\text{E}}^2 }  \right),
\end{equation}
denote the achievable rates of the legitimate link and the eavesdropper link, respectively. Note that in order to maximize the secrecy rate $R_{\text{sec}}$ with given AP transmit beamforming vector $\mathbf{w}$, the design of the IRS reflect beamforming vector $\mathbf{q}$ in general needs to achieve the following two goals. On one hand, the reflected channel $\mathbf{h}_{\text{IU}} \mathbf{Q} \mathbf{H}_{\text{AI}}$ is aligned in phase with the direct channel $\mathbf{h}_{\text{AU}}$ to maximize the received signal power at the user and hence $R_{\text{U}}$. On the other hand, the reflected channel $\mathbf{h}_{\text{IE}} \mathbf{Q} \mathbf{H}_{\text{AI}}$ is in opposite phase with the direct channel $\mathbf{h}_{\text{AE}}$ at the eavesdropper to cancel the signal and thus minimize $R_{\text{E}}$. In general, there is a trade-off in designing $\mathbf{q}$ to achieve the above two goals.

Our objective is thus to maximize the secrecy rate $R_{\text{sec}}$ in \eqref{EquRsec} by jointly optimizing the AP transmit beamforming vector $\mathbf{w}$ and the IRS reflect beamforming vector $\mathbf{q}$. The considered problem is formulated as follows
\begin{subequations}  \label{EquOriProb}
\begin{align}
    \max_{\mathbf{w}, \mathbf{q}} & \; \log_2 \left( 1 + \frac{ | ( \mathbf{h}_{\text{IU}} \mathbf{Q} \mathbf{H}_{\text{AI}} + \mathbf{h}_{\text{AU}} ) \mathbf{w} |^2 }{  \sigma_{\text{U}}^2 }  \right)  \nonumber \\
    & - \log_2 \left( 1 + \frac{ | ( \mathbf{h}_{\text{IE}} \mathbf{Q} \mathbf{H}_{\text{AI}} + \mathbf{h}_{\text{AE}} ) \mathbf{w} |^2 }{ \sigma_{\text{E}}^2 }  \right)  \label{EquOriProbObj}  \\
    \text{s.t.} & \; \| \mathbf{w} \|^2  \leq P_{\text{AP}}  \label{EquConw}  \\
    & \;  | q_n |= 1, \; \forall n.  \label{EquConQ1}
\end{align}
\end{subequations}
Note that in the objective function \eqref{EquOriProbObj}, we have omitted the operator $[\cdot]^+$ without loss of optimality, because the optimal value of our problem must be non-negative\footnote{This fact can be proved by contradiction. If $R_{\text{U}} - R_{\text{E}} < 0$, we can increase its value to zero by setting $\mathbf{w}=\mathbf{0}$ without violating the constraints.}. However, it is still difficult to obtain the optimal solution to problem \eqref{EquOriProb}, since its objective function \eqref{EquOriProbObj} is non-concave with respect to either $\mathbf{w}$ or $\mathbf{q}$, and the unit-norm constraints in \eqref{EquConQ1} are non-convex. Thus, we propose an efficient algorithm for solving problem \eqref{EquOriProb} approximately in the following section.

\vspace{-0.3cm}

\section{Proposed Algorithm for Problem \eqref{EquOriProb}}
In problem \eqref{EquOriProb}, it is observed that the constraint \eqref{EquConw} only contains the variable $\mathbf{w}$ and the constraint \eqref{EquConQ1} only contains the variable $\mathbf{q}$. This motivates us to solve problem \eqref{EquOriProb} by optimizing $\mathbf{w}$ and $\mathbf{q}$ alternately. Specifically, we solve it by solving the following two sub-problems iteratively: one (denoted by sub-problem 1) optimizes $\mathbf{w}$ with given $\mathbf{q}$, and the other (denoted by sub-problem 2) optimizes $\mathbf{q}$ with given $\mathbf{w}$, as detailed in the following two subsections, respectively. Then, we present the overall algorithm and show its convergence and complexity.

\vspace{-0.4cm}

\subsection{Sub-Problem 1: Optimizing $\mathbf{w}$ with Given $\mathbf{q}$}  \label{SecW}
By letting
\begin{align}
    \mathbf{A} & = \frac{1}{\sigma_{\text{U}}^2} ( \mathbf{h}_{\text{IU}} \mathbf{Q} \mathbf{H}_{\text{AI}} + \mathbf{h}_{\text{AU}} )^H ( \mathbf{h}_{\text{IU}} \mathbf{Q} \mathbf{H}_{\text{AI}} + \mathbf{h}_{\text{AU}} ), \\
    \mathbf{B} & = \frac{1}{\sigma_{\text{E}}^2} ( \mathbf{h}_{\text{IE}} \mathbf{Q} \mathbf{H}_{\text{AI}} + \mathbf{h}_{\text{AE}} )^H ( \mathbf{h}_{\text{IE}} \mathbf{Q} \mathbf{H}_{\text{AI}} + \mathbf{h}_{\text{AE}} ),
\end{align}
where the superscript $H$ denotes the conjugate transpose operation, we formulate sub-problem 1 in the following form
\begin{subequations}  \label{EquSubProb1QP}
\begin{align}
    \max_{ \mathbf{w}  } & \; \frac{ \mathbf{w}^H \mathbf{A} \mathbf{w}  + 1 }{ \mathbf{w}^H \mathbf{B} \mathbf{w} + 1  }   \\
    \text{s.t.} & \;  \mathbf{w}^H \mathbf{w}   \leq P_{\text{AP}}.
\end{align}
\end{subequations}
The optimal solution to problem \eqref{EquSubProb1QP} is \cite{Khisti2010}
\begin{equation}  \label{EquOptw}
  \mathbf{w}_{\text{opt}} = \sqrt{P_{\text{AP}}} \mathbf{u}_{\max},
\end{equation}
where $\mathbf{u}_{\max}$ is the normalized eigenvector corresponding to the largest eigenvalue of the matrix $( \mathbf{B} + \frac{1}{P_{\text{AP}}} \mathbf{I}_M )^{-1} ( \mathbf{A} + \frac{1}{P_{\text{AP}}} \mathbf{I}_M )$. Here, $\mathbf{I}_n$ denotes an $n \times n$ identity matrix.

\vspace{-0.4cm}

\subsection{Sub-Problem 2: Optimizing $\mathbf{q}$ with Given $\mathbf{w}$}  \label{SecQ}
Sub-problem 2 can be formulated as
\begin{subequations}  \label{EquSubProb2}
\begin{align}
    \max_{ \mathbf{q} } & \; \frac{ \frac{1}{\sigma_{\text{U}}^2} | ( \mathbf{h}_{\text{IU}} \mathbf{Q} \mathbf{H}_{\text{AI}} + \mathbf{h}_{\text{AU}} ) \mathbf{w} |^2  + 1  }{  \frac{1}{\sigma_{\text{E}}^2} | ( \mathbf{h}_{\text{IE}} \mathbf{Q} \mathbf{H}_{\text{AI}} + \mathbf{h}_{\text{AE}} ) \mathbf{w} |^2  + 1  }      \\
    \text{s.t.} & \; \eqref{EquConQ1}.
\end{align}
\end{subequations}
Since
\begin{equation*}
\mathbf{h}_{\text{IU}} \mathbf{Q} \mathbf{H}_{\text{AI}} = \mathbf{q}^T \text{diag}( \mathbf{h}_{\text{IU}} ) \mathbf{H}_{\text{AI}},
\end{equation*}
\begin{equation*}
\mathbf{h}_{\text{IE}} \mathbf{Q} \mathbf{H}_{\text{AI}} = \mathbf{q}^T \text{diag}( \mathbf{h}_{\text{IE}} ) \mathbf{H}_{\text{AI}},
\end{equation*}
the following equalities hold \cite{Wu2018}
\begin{equation} \label{Equ1}
 \frac{1}{\sigma_{\text{U}}^2}  | ( \mathbf{h}_{\text{IU}} \mathbf{Q} \mathbf{H}_{\text{AI}} + \mathbf{h}_{\text{AU}} ) \mathbf{w} |^2  = \mathbf{s}^H \mathbf{G}_{\text{U}} \mathbf{s} + h_{\text{U}},
\end{equation}
\begin{equation}   \label{Equ2}
\frac{1}{\sigma_{\text{E}}^2} | ( \mathbf{h}_{\text{IE}} \mathbf{Q} \mathbf{H}_{\text{AI}} + \mathbf{h}_{\text{AE}} ) \mathbf{w} |^2  = \mathbf{s}^H \mathbf{G}_{\text{E}} \mathbf{s} + h_{\text{E}},
\end{equation}
where $\mathbf{s} = [ \mathbf{q}^T, 1 ]^T$ and
\begin{equation}
h_{\text{U}} = \frac{\mathbf{h}_{\text{AU}}^* \mathbf{w}^*  \mathbf{w}^T \mathbf{h}_{\text{AU}}^T}{\sigma_{\text{U}}^2} , \; h_{\text{E}} = \frac{\mathbf{h}_{\text{AE}}^* \mathbf{w}^*  \mathbf{w}^T \mathbf{h}_{\text{AE}}^T}{\sigma_{\text{E}}^2}  ,
\end{equation}
and $\mathbf{G}_{\text{U}}$ and $\mathbf{G}_{\text{E}}$ are given in \eqref{EquGU} and \eqref{EquGE}, respectively, shown at the top of next page. Here, the superscript $*$ denotes the conjugate operation.
\begin{figure*}[!t]
	\normalsize
	\begin{equation}  \label{EquGU}
	\mathbf{G}_{\text{U}} = \frac{1}{\sigma_{\text{U}}^2} \begin{bmatrix}
	\text{diag}( \mathbf{h}_{\text{IU}}^* ) \mathbf{H}_{\text{AI}}^* \mathbf{w}^* \mathbf{w}^T  \mathbf{H}_{\text{AI}}^T  \text{diag}( \mathbf{h}_{\text{IU}}  )    &     \text{diag}( \mathbf{h}_{\text{IU}}^* ) \mathbf{H}_{\text{AI}}^* \mathbf{w}^* \mathbf{w}^T \mathbf{h}_{\text{AU}}^T  \\
	\mathbf{h}_{\text{AU}}^* \mathbf{w}^* \mathbf{w}^T  \mathbf{H}_{\text{AI}}^T  \text{diag}( \mathbf{h}_{\text{IU}}  )   &  0
	\end{bmatrix}
	\end{equation}
	\begin{equation}  \label{EquGE}
	\mathbf{G}_{\text{E}} = \frac{1}{\sigma_{\text{E}}^2} \begin{bmatrix}
	\text{diag}( \mathbf{h}_{\text{IE}}^* ) \mathbf{H}_{\text{AI}}^* \mathbf{w}^* \mathbf{w}^T  \mathbf{H}_{\text{AI}}^T  \text{diag}( \mathbf{h}_{\text{IE}}  )    &    \text{diag}( \mathbf{h}_{\text{IE}}^* ) \mathbf{H}_{\text{AI}}^* \mathbf{w}^* \mathbf{w}^T \mathbf{h}_{\text{AE}}^T  \\
	\mathbf{h}_{\text{AE}}^* \mathbf{w}^* \mathbf{w}^T  \mathbf{H}_{\text{AI}}^T  \text{diag}( \mathbf{h}_{\text{IE}}  )   &  0
	\end{bmatrix}
	\end{equation}	
	\hrulefill \vspace{-0.5cm}
\end{figure*}
By substituting \eqref{Equ1} and \eqref{Equ2} into problem \eqref{EquSubProb2}, we rewrite it into a more tractable form as
\begin{subequations} \label{EquQPs}
	\begin{align}
	\max_{ \mathbf{s} } & \; \frac{ \mathbf{s}^H \mathbf{G}_{\text{U}} \mathbf{s} + h_{\text{U}} + 1 }{ \mathbf{s}^H \mathbf{G}_{\text{E}} \mathbf{s} + h_{\text{E}} + 1 } \label{EquQPsObj} \\
	\text{s.t.} & \; \mathbf{s}^H \mathbf{E}_n \mathbf{s} = 1, \forall n,  \label{EquQuadCon}
	\end{align}
\end{subequations}
where the $(i,j)$th element of $\mathbf{E}_n$, denoted by $[\mathbf{E}_n]_{i,j}$, satisfies
\begin{equation}
[\mathbf{E}_n]_{i,j} = \begin{cases}
1 & i=j=n  \\  0 & \text{otherwise}.
\end{cases}
\end{equation}
It is still difficult to find the optimal solution to problem \eqref{EquQPs}, since \eqref{EquQuadCon} for each $n$ is a non-convex quadratic equality constraint, and the objective function \eqref{EquQPsObj} is fractional and non-concave with respect to $\mathbf{s}$. We thus propose an efficient way to find an approximate solution to problem \eqref{EquQPs} as follows.

Let $\text{tr}(\mathbf{Z})$ and $\text{rank}(\mathbf{Z})$ denote the trace and the rank of a matrix $\mathbf{Z}$, respectively. First, we apply the SDR technique \cite{Liu2014} to overcome the non-convexity of \eqref{EquQuadCon}. With $\mathbf{S} \triangleq \mathbf{s} \mathbf{s}^H$, we then re-express problem \eqref{EquQPs} into its relaxed form by dropping the constraint of $\text{rank}(\mathbf{S})=1$ as follows
\begin{subequations}  \label{EquProbSDR}
	\begin{align}
	\max_{ \mathbf{S} \succeq \mathbf{0} } & \; \frac{ \text{tr}( \mathbf{G}_{\text{U}} \mathbf{S} ) + h_{\text{U}} + 1 }{ \text{tr}( \mathbf{G}_{\text{E}} \mathbf{S} ) + h_{\text{E}} + 1 } \\
	\text{s.t.} & \; \text{tr}( \mathbf{E}_n \mathbf{S} ) = 1, \forall n.
	\end{align}
\end{subequations}
Next, we apply the Charnes-Cooper transformation \cite{Liu2014}, i.e., we set $\mu = 1/ [ \text{tr}( \mathbf{G}_{\text{E}} \mathbf{S} ) + h_{\text{E}} + 1  ]$ and $\mathbf{X}=\mu \mathbf{S}$, and transform problem \eqref{EquProbSDR} into the following equivalent non-fractional form  \vspace{-0.3cm}
\begin{subequations}  \label{EquSubProb2SDP}
	\begin{align}
	\max_{ \mu \geq 0 , \mathbf{X} \succeq \mathbf{0} } & \; \text{tr}( \mathbf{G}_{\text{U}} \mathbf{X} ) + \mu (h_{\text{U}} + 1)  \\
	\text{s.t.} \quad & \; \text{tr}( \mathbf{G}_{\text{E}} \mathbf{X} ) + \mu (h_{\text{E}} + 1) = 1  \\
	& \; \text{tr}( \mathbf{E}_n \mathbf{X} ) = \mu, \forall n.
	\end{align}
\end{subequations}
Problem \eqref{EquSubProb2SDP} is a convex semidefinite programming (SDP) problem and thus can be optimally solved by using e.g. the interior-point method \cite{Polik2010}. Finally, to address the omitted constraint $\text{rank}(\mathbf{S})=1$, we apply the standard Gaussian randomization method to obtain an approximate solution to problem \eqref{EquSubProb2}, for which the detail is similar to that in \cite{Wu2018} and thus omitted here for brevity.

\vspace{-0.4cm}

\subsection{Overall Algorithm}
\begin{algorithm}[!t]
	\caption{Proposed Algorithm for Problem \eqref{EquOriProb}.}
	\begin{algorithmic}[1]   \label{Alg}
		\STATE \textbf{Initialization:} Set $k=0$; $\mathbf{w}^{(0)} = \mathbf{h}_{\text{AI}}^H / \| \mathbf{h}_{\text{AI}} \|$; $\mathbf{q}^{(0)}=\mathbf{1}_N$; and $R^{(0)} = f( \mathbf{w}^{(0)} , \mathbf{q}^{(0)}  )$.
		\REPEAT
		\STATE Set $k = k+1$.
		\STATE With given $\mathbf{q}^{(k-1)}$, find the normalized eigenvector corresponding to the largest eigenvalue of $( \mathbf{B} + \frac{1}{P_{\text{AP}}} \mathbf{I}_M )^{-1} ( \mathbf{A} + \frac{1}{P_{\text{AP}}} \mathbf{I}_M )$, and denote it by $\mathbf{u}_{\max}$. Set $\mathbf{w}^{(k)} = \sqrt{P_{\text{AP}}} \mathbf{u}_{\max}$.
        \STATE With given $\mathbf{w}^{(k)}$, solve problem \eqref{EquSubProb2SDP} and apply Gaussian randomization over its solution to obtain an approximate solution $\mathbf{q}^{(k)}$.
        \STATE Set $R^{(k)} = f ( \mathbf{w}^{(k)} , \mathbf{q}^{(k)} )$.
		\UNTIL {$\frac{R^{(k)} - R^{(k-1)}}{R^{(k)}} < \epsilon$.}
	\end{algorithmic}  
\end{algorithm} 

The overall algorithm for problem \eqref{EquOriProb} is presented in Algorithm \ref{Alg}, where $\mathbf{h}_{\text{AI}}$ denotes any row in $\mathbf{H}_{\text{AI}}$, $\mathbf{1}_N$ denotes an $N \times 1$ vector whose elements are all 1, $R^{(k)} = f( \mathbf{w}^{(k)}, \mathbf{q}^{(k)} )$ denotes the objective value of problem \eqref{EquOriProb} with variables $\mathbf{w}^{(k)}$ and $\mathbf{q}^{(k)}$ in iteration $k$, and $\epsilon >0$ denotes a small threshold. Since $R^{(k)}$ is non-decreasing over iterations and is bounded from above, Algorithm 1 is guaranteed to converge. In addition, the complexity of Algorithm \ref{Alg} is mainly due to steps 4 and 5, for which the complexities are $\mathcal{O}(M^{3})$ and $\mathcal{O}((N+1)^{3.5})$ \cite{Polik2010}, respectively. As a result, the complexity of Algorithm \ref{Alg} is $\mathcal{O}( N_{\text{ite}} ( M^{3} + (N+1)^{3.5} ) )$, where $N_{\text{ite}}$ denotes the iteration number of the algorithm and is usually less than 10 for an accuracy of $\epsilon = 10^{-3}$ based on our simulations.

 \vspace{-0.3cm}

\section{Simulation Results}
We provide numerical examples to verify the performance of the proposed ``alternating optimization'' based joint active and passive beamforming design, as compared to the following benchmark schemes:
\begin{itemize}
	\item AP MRT with IRS: it firstly performs maximum ratio transmission (MRT) based beamforming towards the IRS at the AP, i.e., $\mathbf{w} = \sqrt{P_{\text{AP}}} \mathbf{h}_{\text{AI}}^H / \| \mathbf{h}_{\text{AI}} \|$, then designs $\mathbf{q}$ by using step 5 of Algorithm 1.	
	\item Without IRS: it does not use IRS and designs $\mathbf{w}$ according to \eqref{EquOptw} with $\mathbf{q}=\mathbf{0}$. 	
    \item Upper bound: it uses the objective value of the relaxed problem \eqref{EquSubProb2SDP} to obtain a secrecy rate upper bound of the proposed algorithm.
\end{itemize}

We set $M=4$ and assume that the AP, the user, the eavesdropper, and the IRS are located at $(0,0)$, $(150,0)$, $(145,0)$, and $(145,5)$ in meter (m) in a two-dimensional plane, respectively. Under the above setting, since the AP, eavesdropper and user lie on a line, the multiple-input single-output (MISO) channels from the AP to the user and to the eavesdropper are spatially correlated. Specifically, their channel coefficients $\mathbf{h}_{\text{AU}}$ and $\mathbf{h}_{\text{AE}}$ are generated by $\mathbf{h}_{\text{AU}} = \sqrt{ \zeta_0 ( d_0 / d_{\text{AU}} )^{ \alpha_{\text{AU}}  } } \mathbf{g}_{\text{AU}}$ and $\mathbf{h}_{\text{AE}} = \sqrt{ \zeta_0 ( d_0 / d_{\text{AE}} )^{ \alpha_{\text{AE}}  } } \mathbf{g}_{\text{AE}}$, respectively, where $\zeta_0=-30$ dB denotes the path loss at the reference distance $d_0 = 1$ m, $d_{\text{AE}}$ and $d_{\text{AU}}$ denote the distances from the AP to the user and to the eavesdropper, respectively, $\alpha_{\text{AU}}=3$ and $\alpha_{\text{AE}}=3$ denote their path loss exponents, and $\mathbf{g}_{\text{AU}}$ and $\mathbf{g}_{\text{AE}}$ denote the small-scale fading components of $\mathbf{h}_{\text{AU}}$ and $\mathbf{h}_{\text{AE}}$, respectively. We model $\mathbf{g}_{\text{AU}}$ and $\mathbf{g}_{\text{AE}}$ by the spatially correlated Rician fading model with Rician factors $K_{\text{AU}}=K_{\text{AE}}=1$ and spatial correlation matrix $\mathbf{R}$, where $[\mathbf{R}]_{i,j} = r^{|i-j|}$ with $r=0.95$ \cite{Chiani2003}. The other channels $\mathbf{H}_{\text{AI}}$, $\mathbf{h}_{\text{IU}}$ and $\mathbf{h}_{\text{IE}}$ are modeled as independent Rician fading with the corresponding path loss exponents: $\alpha_{\text{AI}}=2.2$, $\alpha_{\text{IU}}=3$ and $\alpha_{\text{IE}}=3$, and Rician factors: $K_{\text{AI}}=K_{\text{IU}}=K_{\text{IE}}=1$. The other parameters are set as $\sigma_{\text{U}}^2 = \sigma_{\text{E}}^2 = -80$ dBm and $\epsilon = 10^{-3}$. The following simulation results are averaged over 1000 random fading realizations.

\begin{figure}[!t]
	\centering
	\includegraphics[width=0.35\textwidth]{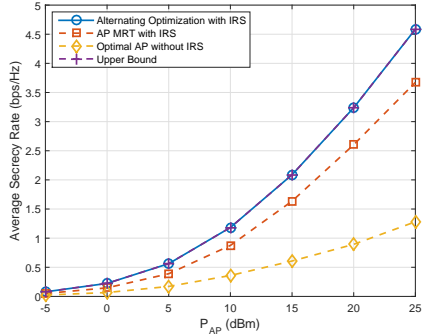}
	\caption{Secrecy rate versus $P_{\text{AP}}$ ($N=64$).}
	\label{FigSRvsP}  \vspace{-0.65cm}
\end{figure}

Fig. \ref{FigSRvsP} shows the average secrecy rates of different schemes versus the AP's transmit power, $P_{\text{AP}}$, when $N=64$. It is observed that the secrecy rate without using the IRS increases slowly with growing $P_{\text{AP}}$. This is because the channels from the AP to the user and to the eavesdropper are highly correlated and the eavesdropper is closer to the AP than the user, thus the channel power of the former is smaller than that of the latter with high probability. As a result, AP transmit beamforming alone can only achieve very limited secrecy rate. In contrast, it is observed that the secrecy rates of the two schemes with the IRS increase significantly with $P_{\text{AP}}$. This is because via optimizing the phase shifts of the reflecting units of the IRS, the reflected signal by the IRS and the direct (non-reflected) signal can be added constructively at the user but destructively at the eavesdropper, thus providing a new degree of freedom to enhance secrecy communication rate. Furthermore, it is observed that the secrecy rate of the proposed joint beamforming design is very close to its upper bound and is significantly higher than that of the heuristic ``AP MRT with IRS'' scheme. This is because in this benchmark scheme, the AP beams towards the IRS only, which cannot fully exploit the aforementioned power enhancement and interference cancellation gains at the user and the eavesdropper, respectively. Thus, the joint design of AP transmit beamforming and IRS reflect beamforming is essential to fully reap the above gains.

\begin{figure}[!t]
	\centering
	\includegraphics[width=0.35\textwidth]{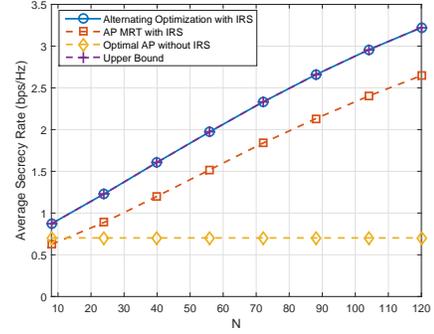}
	\caption{Secrecy rate versus $N$ ($P_{\text{AP}}=15$ dBm).}
	\label{FigSRvsN}  \vspace{-0.65cm}
\end{figure}

Fig. \ref{FigSRvsN} shows the average secrecy rates of different schemes versus the number of IRS's reflecting units, $N$, when $P_{\text{AP}}=15$ dBm. It is observed that the secrecy rate of the proposed algorithm is higher than that of the ``AP MRT with IRS'' scheme, and they both increase as $N$ grows. While their performance gap also increases with $N$, which indicates that with more reflecting elements, the joint transmit and reflect beamforming design becomes more flexible and thus achieves higher gains.

\vspace{-0.35cm}

\section{Conclusion}
This letter proposed a new IRS-aided secure communication system, and investigated the joint optimization of active transmit and passive reflect beamforming to maximize the secrecy rate. By considering a challenging setup under which the conventional system without using IRS has very limited secrecy rate, the proposed system with IRS was shown by simulation able to improve the performance significantly, by exploiting the IRS-enabled power enhancement and interference suppression at the legitimate receiver and the eavesdropper, respectively.


\vspace{-0.4cm}

\begin{thebibliography}{1}
\bibitem{Wu2019}
Q. Wu and R. Zhang, ``Towards smart and reconfigurable environment: Intelligent reflecting surface aided wireless networks,'' [Online]. Available: https://arxiv.org/abs/1905.00152.

\bibitem{Cui2014}
T. Cui, M. Qi, X. Wan, J. Zhao, and Q. Cheng, ``Coding metamaterials, digital metamaterials and programmable metamaterials,'' \textit{Light: Science \& Applications}, vol. 3, no. 10, e218, Oct. 2014.

\bibitem{Tan2016}
X. Tan, Z. Sun, J. M. Jornet, and D. Pados, ``Increasing indoor spectrum sharing capacity using smart reflect-array,'' in \textit{Proc. IEEE ICC}, 2016, pp. 1-6.

\bibitem{Wu2018}
Q. Wu and R. Zhang, ``Intelligent reflecting surface enhanced wireless network via joint active and passive beamforming,'' [Online]. Available: https://arxiv.org/abs/1810.03961.

\bibitem{Wu2018a}
Q. Wu and R. Zhang, ``Intelligent reflecting surface enhanced wireless network: joint active and passive beamforming design,'' in \textit{Proc. IEEE GLOBECOM}, 2018, pp. 1-6.

\bibitem{Wu2018b}
Q. Wu and R. Zhang, ``Beamforming optimization for intelligent reflecting surface with discrete phase shifts,'' in \textit{Proc. IEEE ICASSP}, 2019, pp. 7830-7833.

\bibitem{Huang2018}
C. Huang, A. Zappone, G. C. Alexandropoulos, M. Debbah, and C. Yuen, ``Reconfigurable intelligent surfaces for energy efficiency in wireless communication'', [Online]. Available: https://arxiv.org/abs/1810.06934.

\bibitem{Hu2018}
S. Hu, F. Rusek, and O. Edfors, ``Beyond massive MIMO: the potential of data transmission with large intelligent surfaces,'' \textit{IEEE Trans. Signal Process.}, vol. 66, no. 10, pp. 2746-2758, May 2018.

\bibitem{Khisti2010}
A. Khisti and G. W. Wornell, ``Secure transmission with multiple antennas I: the MISOME wiretap channel,'' \textit{IEEE Trans. Inf. Theory}, vol. 56, no. 7, pp. 3088-3104, Jul. 2010.

\bibitem{Mukherjee2014}
A. Mukherjee, S. A. A. Fakoorian, J. Huang, and A. L. Swindlehurst, ``Principles of physical layer security in multiuser wireless networks: a survey,'' \textit{IEEE Commun. Surveys Tuts.}, vol. 16, no. 3, pp. 1550-1573, Third Quarter 2014.

\bibitem{Liu2014}
L. Liu, R. Zhang, and K.-C. Chua, ``Secrecy wireless information and power transfer with MISO beamforming,'' \textit{IEEE Trans. Signal Process.}, vol. 62, no. 7, pp. 1850-1863, Apr. 2014.

\bibitem{Polik2010}
I. Polik and T. Terlaky, ``Interior point methods for nonlinear optimization,'' in \textit{Nonlinear Optimization}, G. Di Pillo and F. Schoen, Eds., 1st ed. New York, NY, USA: Springer, 2010, ch. 4.

\bibitem{Chiani2003}
M. Chiani, M. Z. Win, and A. Zanella, ``On the capacity of spatially correlated MIMO Rayleigh-fading channels,'' \textit{IEEE Trans. Inf. Theory}, vol. 49, no. 10, pp. 2363-2371, Oct. 2003.


\end{thebibliography}
\end{document}